\documentclass{aa}  

\usepackage{graphicx}
\usepackage{txfonts}
\begin{document} 
%
\newcommand{\herschel}{\emph{Herschel}}
\newcommand{\spitzer}{\emph{Spitzer}}
\newcommand{\planck}{\emph{Planck}}
\newcommand{\akari}{\emph{Akari}}
\newcommand{\swift}{\emph{Swift}}
\newcommand{\iras}{IRAS}
\newcommand{\iso}{ISO}
\newcommand{\alma}{ALMA}
\newcommand{\galex}{GALEX}
\newcommand{\wise}{WISE}
\newcommand{\twomass}{2MASS}
\newcommand{\lsun}{\mbox{L$_\odot$}}
\newcommand{\msun}{\mbox{M$_\odot$}}
\newcommand{\msunyr}{\mbox{M$_\odot$~yr$^{-1}$}}
\newcommand{\rsun}{\mbox{R$_\odot$}}
\newcommand{\lbat}{$L_{\rm 14-195keV}$} 
\newcommand{\lbol}{$L_{\rm bol}$} 
\newcommand{\lbolagn}{$L_{\rm bol,AGN}$} 
\newcommand{\ltorus}{$L_{\rm Torus}$} 
\newcommand{\lfir}{$L_{\rm FIR}$} 
\newcommand{\lir}{$L_{\rm IR}$} 
\newcommand{\reblue}{$R_{\rm e,70}$} 
\newcommand{\regreen}{$R_{\rm e,100}$} 
\newcommand{\rered}{$R_{\rm e,160}$} 
\newcommand{\sigfir}{$\Sigma_{\rm FIR}$} 
\newcommand{\sigsfr}{$\Sigma_{\rm SFR}$} 
\newcommand{\deltare}{$\Delta_{\rm Re}$} 
\newcommand{\tdust}{$T_{\rm dust}$}
\newcommand{\mstellar}{$M_{\ast}$}     
\newcommand{\ergs}{erg~s$^{-1}$}    
\newcommand{\ergscm}{erg~s$^{-1}$~cm$^{-2}$} 
\newcommand{\mum}{\mbox{$\mu$m}}
\newcommand{\cii}{[C{\sc ii}]}

   \title{Local \swift\/-BAT active galactic nuclei prefer circumnuclear star formation}


   \author{D.~Lutz\inst{1}
          \and T.~Shimizu\inst{1}
          \and R.I.~Davies\inst{1}
          \and R.~Herrera-Camus\inst{1}
          \and E.~Sturm\inst{1}
          \and L.J.~Tacconi\inst{1}
          \and S.~Veilleux\inst{2}
          }

   \institute{Max-Planck-Institut f\"ur extraterrestrische Physik,
              Giessenbachstra\ss\/e, 85748 Garching, Germany\\
              \email{lutz@mpe.mpg.de}
        \and
             Department of Astronomy, University of Maryland, 
             College Park, MD 20742, USA
             }

   \date{Received 23 June 2017; Accepted 1 Sept 2017}

  \abstract{
We use \herschel\ data to analyze the size of the far-infrared 70~\mum\ 
emission for 
z$<$0.06 local samples of 277 hosts of \swift\/-BAT selected active galactic 
nuclei (AGN), and 515 
comparison galaxies that are not detected by BAT. For modest far-infrared 
luminosities 
8.5$<$log(\lfir\ [\lsun\/])$<$10.5, we find large scatter of half light 
radii \reblue\
for both populations, but a typical \reblue\/$\lesssim$1~kpc for the BAT 
hosts that is only half that of comparison galaxies of same 
far-infrared luminosity. The result mostly reflects 
a more compact distribution of star formation (and hence gas) in the AGN 
hosts, but compact
AGN heated dust may contribute in some extremely AGN-dominated systems.
Our findings are in support of an AGN--host coevolution where accretion onto
the central black hole and star formation are fed from the same gas reservoir,
with more efficient black hole feeding if that reservoir is more concentrated.
The significant scatter in the far-infrared sizes emphasizes that we are 
mostly probing spatial scales much larger than those of actual accretion, 
and that rapid accretion variations can 
smear the distinction between the AGN and comparison categories. Large 
samples are hence needed to detect structural differences that favour 
feeding of the
black hole. No size difference AGN host vs. comparison galaxies is
observed at higher far-infrared luminosities log(\lfir\ [\lsun\/])$>$10.5
(star formation rates $\gtrsim$6~\msunyr\/), possibly because 
these are typically reached in more compact regions in the first place.  
  }


   \maketitle
%

\section{Introduction}

The evolution of galaxies and their central black holes is linked
by the gas supply that is 
needed both for feeding star formation and for 
accretion onto the black hole, and by feedback effects that the black hole 
exerts on the galaxy during its phases of activity. 
But for several related reasons, the `feeding' link between star formation and
black hole accretion cannot be tight, and must be difficult to determine
from individual objects or small samples.
First, the two phenomena occur on very different spatial scales. Stars may 
form in gas present on a wide variety of scales, at distances from
the nucleus between parsecs and 10~kpc or more. In contrast, accretion
onto the black hole directly relates only to the material in its immediate
surroundings, from the black hole sphere of influence down to 
the event horizon.
Second and related to the smaller scales, black hole accretion rates (BHAR)
and AGN luminosities can vary rapidly on timescales of years, much shorter than
the millions of years or longer on which star formation rates (SFR) vary. 
As a consequence, the relation between average SFR and BHAR that is observed 
for large SFR selected samples \citep[e.g.,][]{chen13,delvecchio15} is largely 
erased by the AGN variations when looking at the SFR of samples selected by 
the instantaneous BHAR
\citep[e.g.,][]{shao10,mullaney12,rosario12,stanley15,shimizu17}.
  
A direct approach to the feeding problem would be to map and compare the 
molecular gas distributions in active and inactive galaxies, from global gas 
content down to the smallest scales accessible to current mm interferometers.
This is an expensive project if large samples are needed to probe whether
intriguing phenomena in the gas distribution and kinematics 
\citep[e.g.,][]{garcia-burillo03,garcia-burillo05,haan09} 
are indeed more prevalent in active galaxies and related to AGN feeding. 
Alternatively, star formation can be 
mapped in active and inactive galaxies, using a star formation indicator 
that is little disturbed by the AGN. In this type of study, 
\citet{diamond-stanic12} analysed 84 Seyferts and used circumnuclear star 
formation rates from mid-infrared PAH features and global star formation 
rates from extended mid-infrared continuum to argue for a stronger link 
between BHAR and circumnuclear or kpc scale 
star formation than between BHAR and global SFR.  

The best contrast between the SED of a star forming galaxy and an AGN SED
is reached in the far-infrared, making the far-infrared emission a good tool 
for measuring SFRs of AGN hosts, except for the AGN with the most extreme 
ratio of \lbolagn\ and SFR, for which the contribution of AGN heated dust 
to the far-infrared 
can be significant. The PACS instrument \citep{poglitsch10} on board
\herschel\/\footnote{Herschel is an ESA space observatory with science 
instruments provided by European-led Principal Investigator consortia and 
with important participation from NASA.}
\citep{pilbratt10} has dramatically improved sensitivity and 
spatial resolution of far-infrared photometric mapping. Several studies have
already used \herschel\ to study the far-infrared structure of local 
AGN hosts. 
\citet{mushotzky14} reported a large fraction of almost point like sources 
($\sim$ 1/3 unresolved sources at 70~\mum\/, with a 5.8\arcsec\ PSF), 
small FIR sizes, and 
high surface brightnesses in a large z$<$0.05 SWIFT-BAT selected AGN sample, 
but with very limited comparison to inactive galaxies.    
\citet{garcia-gonzalez16} discuss \herschel\ images of 33 nearby RSA Seyferts,
with nuclei unresolved at FWHM of order a kpc in 1/3 of the 70~\mum\ images 
and most flux emerging within a radius of 2~kpc for 85\% of the galaxies.
\citet{lutz16} find far-infrared sizes of a sample of local PG QSOs 
to be consistent with non-active galaxies of the same FIR luminosity, but the 
modest sample size and significant fraction of upper size limits, due to the 
large distance of the QSOs, are limiting factors. 
Here, we use the \herschel\ archive to assemble large samples of
local AGN and inactive comparisons, and measure and 
compare in a consistent manner the far-infrared sizes of AGN hosts and other 
galaxies.

Section~2 discusses the sample and data analysis, Section~3 reports the 
results, and Section~4 discusses the size differences we find in terms of
the properties of star formation in AGN hosts. 
We adopt an $\Omega_m =0.3$, $\Omega_\Lambda =0.7$ and 
$H_0=70$ km\,s$^{-1}$\,Mpc$^{-1}$ cosmology, redshift-independent distances
from NED, if available, for z$<$0.01 galaxies, a \citet{chabrier03} IMF, 
a conversion
${\rm SFR} = 1.9\times 10^{-10}L_{FIR=40-120\mum}$ as appropriate for the
\citet{kennicutt98} conversion corrected to Chabrier IMF, and a ratio 1.9
of 8--1000~\mum\ IR and 40--120~\mum\ FIR luminosity.

\section{Data}

The AGN sample used in this study is based on the 58 month version of the 
\swift\ Burst Alert Telescope (BAT) AGN 
sample\footnote{\url{https://swift.gsfc.nasa.gov/results/bs58mon/}} 
\citep[see also][]{baumgartner13}. Uniform sky 
coverage and selection in 14-195~keV very hard X-rays, that are detecting all 
but the fully Compton-thick AGN, make this sample an excellent basis for 
studies of local moderate luminosity AGN 
(typical log(\lbolagn\ [\ergs\/])$\sim$44.5) and their hosts. 
The z$<$0.05 objects of this sample have been observed with \herschel\/-PACS 
and SPIRE \cite[see][for selection, observations, and results from these data]
{mushotzky14,melendez14,shimizu15,shimizu16,shimizu17}. Most were observed
in project OT1\_rmushotz\_1 and the rest in a variety of other projects.
We use BAT counterpart identification and X-ray luminosities from these 
references. Where discussing AGN bolometric luminosities, we adopt 
$L_{\rm Bol,AGN} = 10.5\times L_{\rm 14-195keV}$  \citep{melendez14}.

As non-AGN comparison objects we use a broad set of \herschel\/-PACS 
observations of local galaxies. These include the luminous infrared 
galaxies from RBGS \citep{sanders03} 
and the KINGFISH galaxies \citep{kennicutt11} that were already 
used by \citet{lutz16} to study local scaling relations of FIR size and 
surface brightness. In order to improve the number of non-BAT comparison
targets in the relevant range of infrared luminosities and redshifts, we
searched the \herschel\ archive for other approved programmes with 
PACS photometric observations of nearby galaxies and retrieved and processed
the observations actually obtained. We also included galaxies serendipitously
observed in the maps as long as they were clearly identified, separated from 
the original target, and with known redshift. Comparison galaxies may host
low luminosity AGN that are below the BAT detection threshold. We did not 
attempt to identify these by other means. 
Classification as BAT-AGN or comparison solely
depends on detection in the 58 month BAT catalogue and not on \herschel\ 
observing program, i.e. some of the BAT sources were observed by projects
mostly contributing to the comparison sample.

\begin{figure}
\centering
\includegraphics[width=\hsize]{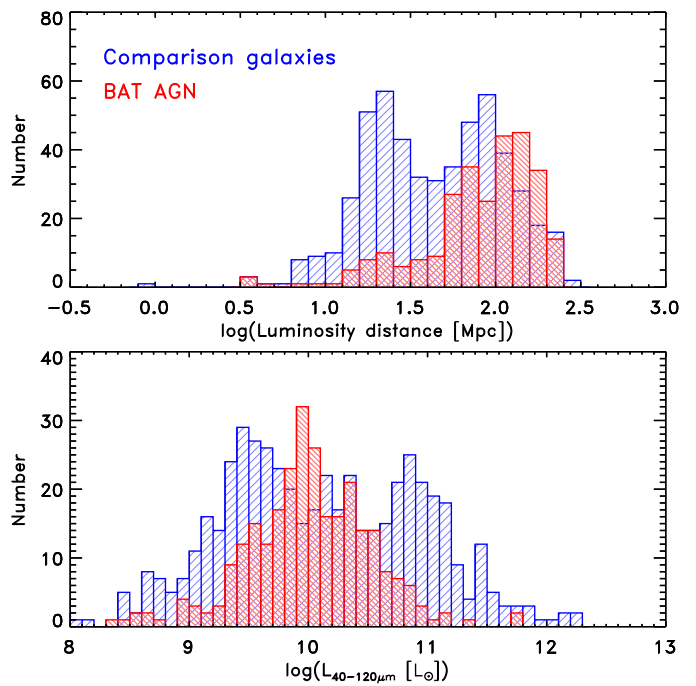}
\caption{Distribution of distances and far-infrared luminosities for
the samples of z$<$0.06 BAT AGN and non-BAT comparison galaxies.}
\label{fig:sample}
\end{figure}

We restrict both the BAT-detected sample and the comparison sample to
z$<$0.06. This excludes a few higher redshift BAT detected AGN and QSOs 
observed in other projects, as well as higher redshift comparison
galaxies. The redshift cut 
maintains the systematic BAT redshift selection used by the above references
and provides comparison galaxies
with similar redshift and scale, avoiding distant and faint targets for which
size information is more difficult to obtain given the \herschel\ beam.   

AGN are hosted by massive galaxies, see for example \citet{kauffmann03} and, 
specifically for BAT AGN, \citet{koss11}. We hence require for our comparison 
galaxies a minimum near-infrared luminosity as a proxy to stellar mass. For 
sensitivity reasons, we extract this from the 3.4~\mum\ band 1 ALLWISE 
catalogs that are based on the \wise\ mission \citep{wright10}, 
and in a few cases our own aperture photometry on \wise\ 
3.4~\mum\ images. Specifically, we require 
$log(\nu L_\nu(3.4)\ [\lsun\/])>8.5$ which is equivalent to 
$log(M*\ [\msun\/])\gtrsim 9.5$ \citep[e.g.][]{wen13}, in good agreement with 
the total mass range of AGN hosts in \citet{kauffmann03} and \citet{koss11}. 
Indeed, only one of the BAT AGN hosts discussed below 
(2MASXi~J1802473-145454, $log(\nu L_\nu(3.4))=8.36$ ) falls slightly below 
this threshold, which may also relate to the lack of a 
redshift-independent distance for this nearby source.
For the comparison galaxies that are already excluding the powerful BAT AGN, 
3.4~\mum\ contamination by AGN hot dust is not significant, and we have 
verified that a cut using 0.5~dex higher K-band luminosity from the \twomass\ large galaxy atlas, extended source catalog,
and point source catalog results in a sample that is very similar
to the \wise\ cut.
 
\begin{figure*}[t]
\centering
\includegraphics[width=\hsize]{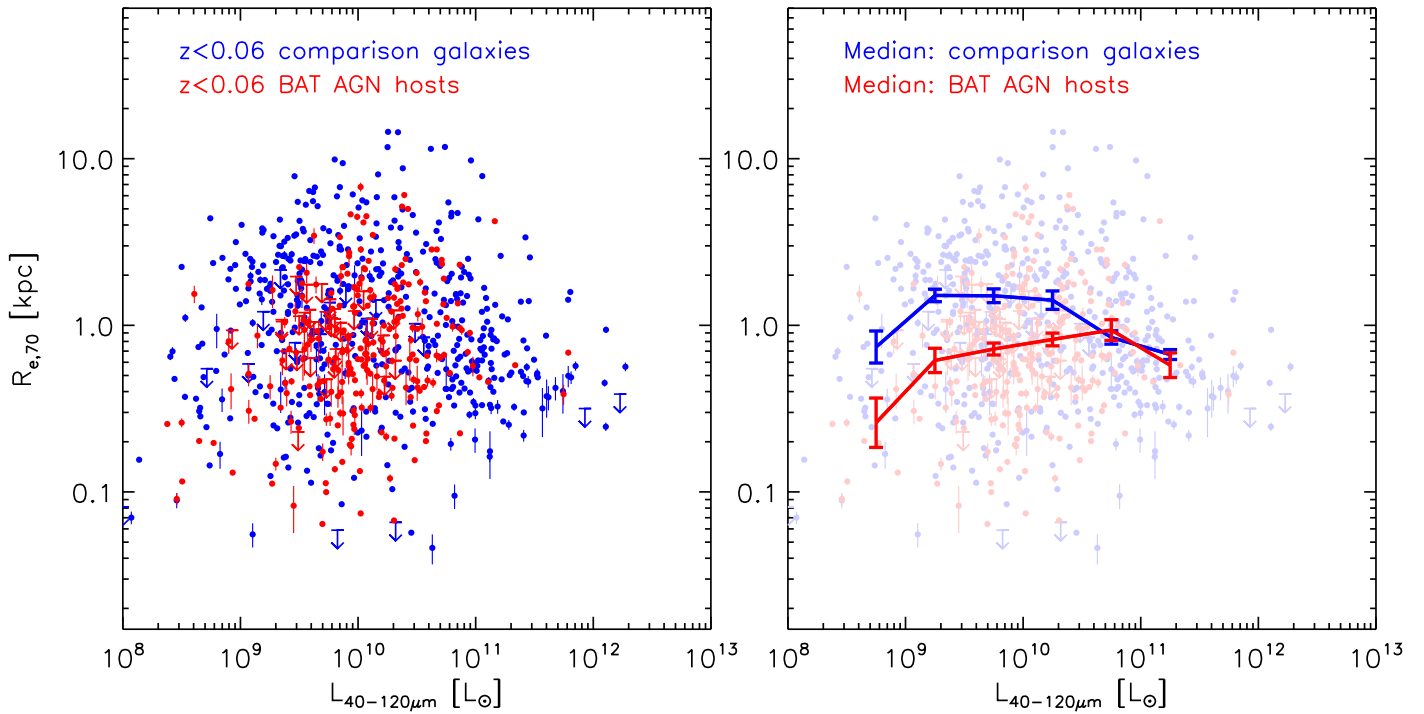}
\caption{Far-infrared 70~\mum\ half light radii as a function of 
FIR luminosity, 
for the Herschel-BAT sample and the comparison sample. 
Left panel: as observed. Right panel: Median values and 
the 1$\sigma$ uncertainties of these medians are overplotted for 0.5~dex 
bins in \lfir\/.}
\label{fig:size_lfir}
\end{figure*}

Reduction of the PACS data uses the methods described in \citet{lutz16}.
Briefly, we use \herschel\ archive data processed with SPG 13.0.0 -- this 
includes the improved gyro-reconstructed pointing history that 
minimizes effects of pointing jitter on the PSF. For objects that are 
unresolved or just resolved by \herschel\/-PACS we use our own dedicated
processing that provides a stable PSF (verified with observations of 
reference stars), at the expense of not preserving large scale emission.
For larger sources we hence use the pipeline JSCANAM \citep{graciacarpio15}
maps. In both cases, we fit sources detected at SNR$>$10 with a 2-dimensional 
gaussian, and derive the source FWHM by subtracting from the circularized 
observed FWHM in quadrature the circularized PSF FWHM that is appropriate for
the source's far-infrared color. Weaker or undetected
sources are not used for our size analysis. This approach can homogeneously
assign a single scale to large samples of \herschel\ observed targets, in the
very highest S/N  cases down to 1/5 of the PSF. Conversely, as discussed
in \citet{lutz16}, the approach provides only a simplifying measure 
of the complex structure of nearby galaxies. For example in case of a 
compact circumnuclear starburst superposed on a large disk, the fit result
may reflect the compact component or the disk, dependent on their relative
fluxes. We re-use the \citet{lutz16} results with the exception of 
some updated
redshift-independent distances and the use of different error maps for fits
to KINGFISH galaxies. While we obtained measurements for all three PACS 
filters if available, we only use below the 70~\mum\ sizes and 
half light radii 
\reblue\ because of the sharpest PSF and closer link to active star 
formation than for longer wavelengths.

Most PACS observations of nearby galaxies have used the combination of the 
70 and 160~\mum\ PACS bands or all three bands including 100~\mum\/. A notable
exception is the Herschel Reference Survey HRS \citep{boselli10}, which used
for its PACS observations only 100 and 160~\mum\ \citep{cortese14}. 
In order to 
preserve the many targets of this program and a few others for the non-BAT 
comparison sample, we have scaled the \regreen\ measured from their data to 
\reblue\ via multiplication by 0.85, which is the median for 
non-BAT objects observed in all bands. We have verified that exclusion of these
objects, while reducing the comparison object statistics at low IR 
luminosities, preserves the immediate result that is reported below in 
Fig.~\ref{fig:size_lfir}.

Our measurements are not meaningful and accurate for pairs or interacting 
systems that have 
a separation of the components of order the PSF width, and down to 
the scales accessible to the method. The measurement then mostly reflects 
distance and relative strength of FIR emission of the two galaxies rather 
than the structure of a galaxy. As described in \citet{lutz16}, we did not fit
intermediate separation doubles that would need more complex reduction
and fitting schemes, and flagged fit results for closer doubles as identified
through multiwavelength data. Neither of the two are used below. BAT AGN
and comparison samples are restricted to single galaxies or components of 
wide doubles that can be analysed separately.

With these restrictions to z$<$0.06 and to single or widely separated
objects, our \herschel\ archival samples of BAT AGN and comparison galaxies
include 277 and 515 galaxies, respectively\footnote{
\herschel\ projects contributing to the BAT and/or comparison samples via
original targets or serendipitously observed galaxies include 
KPGT\_cwilso01\_1, 
KPGT\_esturm\_1,  
KPGT\_smadde01\_1, 
KPOT\_jdavie01\_1, 
KPOT\_rkennicu\_1, 
GT1\_lspinogl\_2,  
GT1\_mbaes\_1,     
GT1\_msanchez\_2,  
OT1\_bholwerd\_1,  
OT1\_dsanders\_1, 
OT1\_lcortese\_1,  
OT1\_lho\_1,       
OT1\_rmushotz\_1,  
OT1\_sveilleu\_1,  
OT1\_vwild\_1,    
OT2\_aalonsoh\_2,  
OT2\_aleroy\_2,    
OT2\_bholwerd\_3,  
OT2\_datlee\_1,    
OT2\_dpisano\_1,  
OT2\_emurph01\_3,  
OT2\_jrigby\_3,    
OT2\_jsmith01\_2,  
OT2\_kwestfal\_2,  
OT2\_lhunt\_4,    
OT2\_mboquien\_3,  
OT2\_mboquien\_4,  
OT2\_mhaynes\_2.  
}. Fig.~\ref{fig:sample} shows their 
distributions in distance and far-infrared luminosity. The comparison sample
covers the range of the BAT AGN. As expected for a purely archival sample, its
detailed distribution differs from the BAT detected sample, partly due to 
the significant numbers 
contributed by dedicated projects aiming, for example,  
at typically distant IR-luminous galaxies or at nearby (often Virgo cluster) 
galaxies. We have tested the effect of randomly eliminating 80\% of the
log(D$_{\rm L} [Mpc])$<1.7 comparison galaxies, a cut that makes the 
BAT and comparison redshift histograms quite similar. While statistical 
errors increase, the basic result of Fig.~\ref{fig:size_lfir} below is again 
preserved. The difference in \lfir\ distributions 
(Fig.~\ref{fig:sample} bottom) is considered by our analysis, because 
comparisons are done as a function of \lfir\/.

Derived half light radii at 70~\mum\ \reblue\/ for our 
samples as well
as far-infrared and BAT luminosites are
listed in Table~\ref{tab:data} in  Appendix~\ref{app:datatable}. Part
of the data are shown for guidance, the full
table is electronically available via the CDS VizieR service.

\section{Results}

Half light radii at 70~\mum\/, \reblue\/, are shown 
in Fig.~\ref{fig:size_lfir} 
as a function of FIR luminosity, for both the z$<$0.06 BAT AGN hosts and the 
z$<$0.06 comparison galaxies. First, as already emphasized by 
\citet{lutz16}, there 
is a large $\sim$2~dex spread of far-infrared size of galaxies at moderate IR 
luminosities - such luminosities can be produced in compact circumnuclear 
regions as well
as by star formation spread over large disks. Fig.~\ref{fig:size_lfir} 
shows that such a large size spread is observed for both the BAT AGN hosts 
and the comparison galaxies, and that the distributions overlap, with an 
indication for smaller sizes of the BAT hosts. 

In order to quantify this size difference, we have computed  
the median far-infrared half light radius for both categories 
and for six bins of 0.5~dex width in logarithm of the far-infrared luminosity.
We derive the median value, median absolute deviation MAD, and uncertainty 
$\sigma = 1.4826\times MAD /\sqrt{N}$, for the half light radius \reblue\ 
of the objects in each bin. These are plotted in the right panel
of Fig.~\ref{fig:size_lfir} and are listed in 
Table~\ref{tab:size_lfir}. Given the large scatter of individual objects, 
the subsample size N in each bin is essential for a meaningful 
comparison. In the 
computation of the median and MAD, we have used the upper size limits 
(12\% of the BAT hosts and 3\% of the comparisons) at their nominal values, 
this is conservatively underestimating differences BAT vs. comparison as 
reported below.

For the modest FIR luminosity bins up to log(\lfir\ [\lsun\/])=10.5, that is
SFR$\lesssim$6~\msunyr\ if the far-infrared is due to star formation, we find AGN hosts
a factor $\sim$2 smaller in the FIR than the comparison galaxies. The 
difference reaches up to the 5.8$\sigma$ level, for the bin centered on 
log(\lfir\/ [\lsun\/])=9.75. No such difference is seen for higher FIR 
luminosity (higher SFR), 
but the statistics for the AGN hosts is more limited in this regime.
This is even more true for the PG QSOs studied by \citet{lutz16}. The size of 
their FIR emission (their Fig.~10 right) is fully consistent with that of the
BAT hosts at same FIR luminosity in Fig.~\ref{fig:size_lfir}. The 
PG QSO statistics is too limited for any meaningful separation into subsamples 
grouped by FIR luminosity, but the general pattern with smallish sizes 
for the very few modest FIR luminosity PG hosts, but sizes close to 
comparison galaxies for the more numerous higher \lfir\ is consistent with 
what is reported here with better statistics for the BAT hosts.  

These results are consistent with the important $\lesssim$2~kpc `point source'
contributions reported for PACS far-infrared images of the BAT sample by 
\citet{mushotzky14}, and extend that work via improved size measurements and a 
systematic comparison to non-BAT galaxies.

To assist the discussion of the size difference between AGN hosts and 
comparison galaxies, we define for each 
BAT AGN host in the range 8.5$\leq$log(\lfir\ [\lsun\/])$\leq$10.5 a 
`size excess' quantity

\begin{equation} 
\Delta R_e = log(R_{e,70,AGN}) - log(Median(R_{e,70,Comparison}))
\label{eqn:sizeexcess}
\end{equation}

where the median is taken over all comparison galaxies in the 0.5~dex wide
\lfir\ bin of the given AGN host. By referencing AGN to the value in 
their bin, we consider the minor trends for the comparison galaxies seen
in Fig.~\ref{fig:size_lfir} but are able to keep the full  
8.5$\leq$log(\lfir\ [\lsun\/])$\leq$10.5 statistics.
   
\begin{table*}
\caption{FIR half light radii of BAT AGN hosts and comparison galaxies}
\begin{tabular}{crrrrrrrr}\hline
Bin center&\multicolumn{4}{c}{BAT AGN hosts}&\multicolumn{4}{c}{Comparison galaxies}\\
log(\lfir\ [\lsun ])&N&\multicolumn{3}{c}{log(\reblue\ [kpc])}&N&\multicolumn{3}{c}{log(\reblue\ [kpc])}\\ 
&&Median&MAD&$\sigma$&&Median&MAD&$\sigma$\\ \hline
 8.75&   9& -0.585&  0.300&  0.148&  29& -0.130&  0.349&  0.096\\
 9.25&  29& -0.211&  0.265&  0.073&  94&  0.179&  0.244&  0.037\\
 9.75&  99& -0.142&  0.240&  0.036& 111&  0.177&  0.294&  0.041\\
10.25&  93& -0.085&  0.248&  0.038&  92&  0.151&  0.357&  0.055\\
10.75&  38& -0.028&  0.257&  0.062&  96& -0.073&  0.265&  0.040\\
11.25&   4& -0.239&  0.100&  0.074&  62& -0.175&  0.158&  0.030\\
\hline
\end{tabular}
\tablefoot{MAD = median absolute deviation. Uncertainty $\sigma = 1.4826\times MAD /\sqrt{N}$ }
\label{tab:size_lfir}
\end{table*}

\begin{figure}
\centering
\includegraphics[width=\hsize]{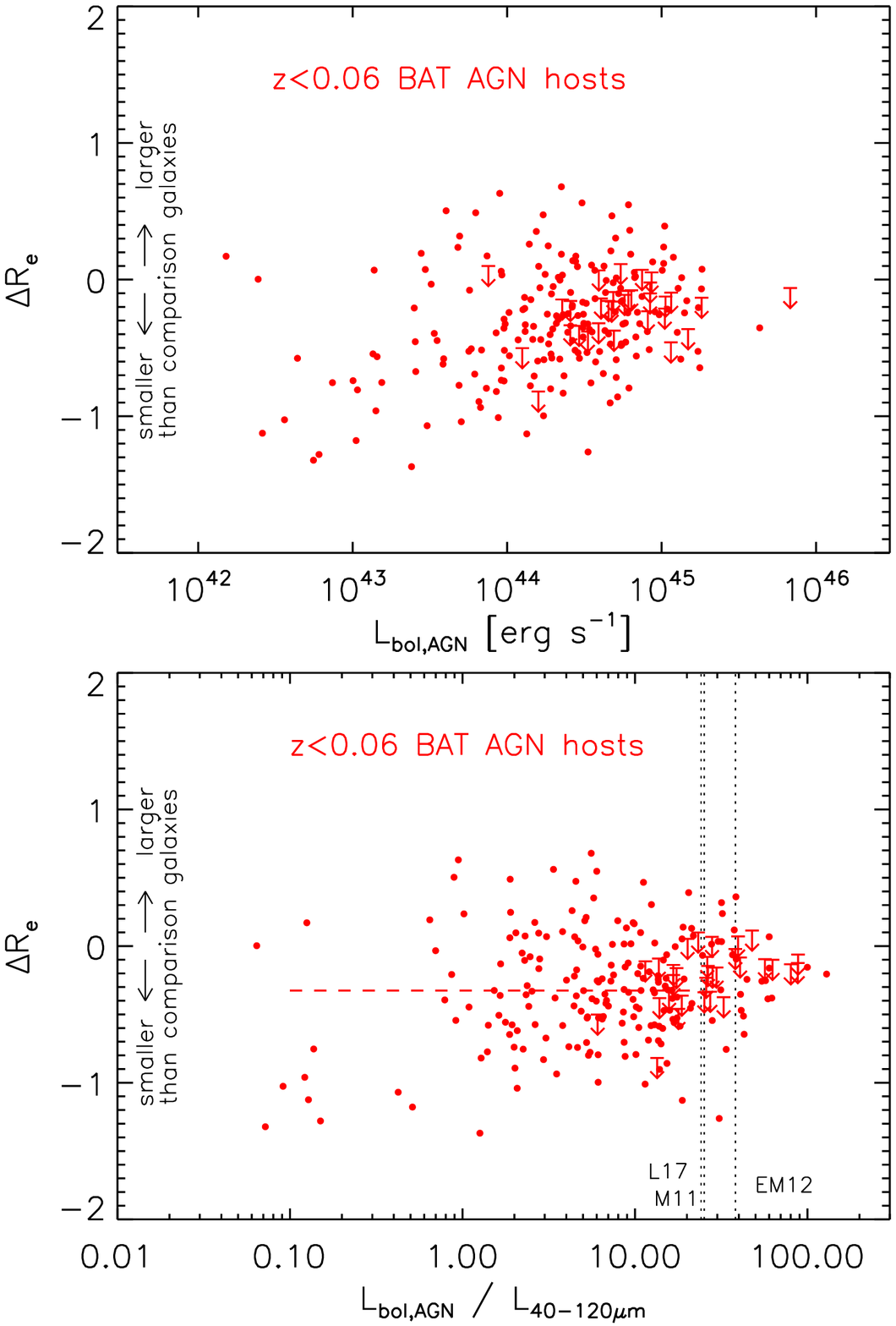}
\caption{`Size excess' $\Delta R_e$ (Eqn.~\ref{eqn:sizeexcess}) of BAT AGN 
hosts compared to the median value for comparison galaxies. 
Top: As a function of AGN bolometric luminosity.
Bottom: As a function of the ratio of AGN bolometric luminosity to 
far-infrared luminosity. Vertical dotted lines mark \lbol\//\lfir for three 
version of pure AGN `intrinsic' SEDs - the extended \citet{mor12} one,  
an SED derived by \citet{lani17}, and a \citet{mullaney11} `Hi. Lum' SED,
extended from 6\mum\ down to 1\mum\/.
 A size deficit for the AGN hosts is observed
also for \lbolagn\//\lfir$<$25 (dashed horizontal median line), where the AGN 
contribution to the FIR should be minor according to these SEDs.
}
\label{fig:lbol_deltare}
\end{figure}

\section{Discussion}

\subsection{Compact star formation or AGN heated dust?}

Two hypotheses are obvious to explain the factor $\sim$2 smaller sizes of the
FIR emission in moderately FIR luminous AGN hosts: First, a more compact gas 
distribution would lead to more compact star formation and FIR emission, and, 
at least in the long term average, boost feeding of the central SMBH, making 
the galaxy more likely to be detected as an AGN. Second, direct heating
of dust by the AGN radiation could lead to a central emission spike that 
is significant not only for the mid-infrared AGN emission from warm dust, but 
even out to the far-infrared. This second effect should be most prominent 
for the most powerful AGN and for those with the largest ratio of \lbolagn\ 
and \lfir\/.

In Fig.~\ref{fig:lbol_deltare}, we plot the `size excess' \deltare\ 
(Eqn.~\ref{eqn:sizeexcess}) for the
BAT AGN as a function of AGN bolometric luminosity, and as a function of the 
ratio of AGN bolometric luminosity and FIR luminosity. No clear trend is seen 
in either diagram, and compact FIR emission (small $\Delta R_e$) is observed
also for BAT systems with small \lbolagn\ and small ratio to \lfir\/.

The lower panel of Fig.~\ref{fig:lbol_deltare} can also be directly compared
to expectations for the `intrinsic' infrared SED of AGN, i.e. the SED of dust
that is directly heated by the AGN, excluding the stellar heated host 
contribution. Such intrinsic SEDs are very difficult to model from first
principles, as long as one makes use of the considerable freedom to modify 
the geometric arrangement
of dust around the AGN.  It is however worth noting that for
radiative transfer models with plausible geometric assumptions 
\citep{schneider15,duras17}, AGN heated emission falls short of fully 
dominating the far-infrared even for some of the most luminous high-z QSOs.
Empirical determinations of the intrinsic
SED, by subtraction of a host SED from the observed total SED, are also
uncertain. Variations among such empirical SEDs arise  from the uncertainty 
in measuring the host SFR (e.g.  via the mid-infrared PAH emission), and in 
adopting the appropriate host SED template for the given SFR. We overplot 
in the lower panel of Fig.~\ref{fig:lbol_deltare} 
expectations for a number of intrinsic SED templates from the literature.
These intrinsic SEDs serve to indicate a range of recent attempts
to determine intrinsic AGN SEDs in the infrared 
\citep[see also][]{netzer07,lyu17a}.
We assume for these \lbolagn = 2~$\times$~\ltorus\/, where \ltorus\ is
taken as the 1--1000~\mum\ integral over the intrinsic AGN SED.  
\citet{mor12} derive an intrinsic SED with relatively low FIR emission, which
\citet{netzer14} extrapolated to longer wavelengths by a modified blackbody.
The intrinsic SED of \citet{mullaney11} (their `Hi. Lum' one which is 
appropriate for our \lbolagn\ range, and which we have amended with 
the \citet{mor12} SED for 
1--6\mum\/) as well as the recent PAH-based SED of \citet{lani17} have stronger
FIR emission.  The yet stronger FIR suggested by \citet{symeonidis16} is 
problematic \citep{lyu17a,lani17} and is not used here. 
The vertical dotted lines
in  Fig.~\ref{fig:lbol_deltare} are based on the simple 40--120~\mum\ integral 
which is about twice $\nu L_{\nu,70}$, with variation between intrinsic AGN
SEDs. They are hence conservatively overestimating the AGN effect at the PACS 
wavelengths proper.   

If one of these intrinsic AGN SEDs were universal and the small FIR sizes of
BAT hosts were due to AGN heated dust, one should expect a downward trend to 
the right of the corresponding line in Fig.~\ref{fig:lbol_deltare} bottom. 
This is not observed. Some effect might still be hidden behind limited 
statistics at very high \lbolagn\//\lfir\ and the fact 
that there will be no strictly universal AGN SED, as also clearly seen from 
slope variations in the mid-infrared \citep[e.g.][]{netzer07,lyu17}. Still,
our results may be slightly in favour of the more FIR-weak intrinsic SEDs 
(more rightward vertical line). More important, the median \deltare\ is 
\mbox{-0.33}
for the AGN hosts with \lbolagn\//\lfir\/$<$25. That means, we find the same
factor $\sim$2 size deficit also for the AGN that are weak compared to their
host, and where AGN heated FIR dust should provide an unimportant 
contribution given the intrinsic AGN SEDs.

We conclude that the FIR size deficit of BAT AGN hosts is mostly a host 
property, reflecting a different distribution of gas and star formation
than in comparison galaxies of same \lfir\/.
A dominant AGN contribution to the FIR emission and its compactness may 
nevertheless be present in some individual BAT sources. This may also be true
for some very FIR-faint BAT AGN that do not enter our analysis because of 
insufficient 70~\mum\ signal-to-noise.

\subsection{Circumnuclear vs. disk scale star formation in AGN hosts}

We have established a tendency for star formation in BAT AGN hosts,
as traced by far-infrared emission, to be spread over a region that is
typically only half the 
size of that in comparison galaxies.
Turned around, this implies that for a given star formation rate, or 
related to it molecular gas mass,
an AGN is more likely to be fuelled if that supply is more centrally 
concentrated. This is completely plausible in an AGN-host coevolution where
black hole and star formation are fed from a common reservoir.   

The large scatter in FIR sizes for both AGN hosts and comparisons, and the 
need for a large sample to establish this size difference are an obvious 
consequence of the diverse star formation morphologies of local galaxies,
and of the different spatial scales of star formation and black hole accretion.
Their link, while present, is not tight. For the same reason, we do not see 
a relation to AGN bolometric luminosity over the range of the BAT sample 
(Fig.~\ref{fig:lbol_deltare}) -- accretion may vary over such a dynamic 
range much more rapidly than SF phenomena will do.

While more concentrated than in non-active galaxies, the star formation in 
the BAT AGN hosts typically does not occur in extreme events. With typical 
star formation rates of $\sim$1~\msunyr\ and half light radii of 
$\lesssim$1kpc, they correspond
to a fairly undramatic local disk galaxy SFR that just is somewhat more 
spatially 
concentrated. Very long periods would be needed to significantly modify 
the stellar population over this spatial scale.

At this point, it is interesting to return to the fact that we observe a
far-infrared size deficit for AGN hosts only up to modest \lfir\/, or 
equivalently SFR $\lesssim$6~\msunyr\
(Fig.~\ref{fig:size_lfir}).
No significant size difference is observed for higher FIR luminosity. With 
the more limited statistics in these bins, a factor $\sim$2 size difference 
(as observed for the lower \lfir\ bins) 
is excluded at the 4.7$\sigma$ level for the bin centered on log(\lfir\/)=10.75
and 2.8$\sigma$ for the bin centered at log(\lfir\/)=11.25. While yet larger 
samples would be desirable to ultimately establish this  different behaviour, 
one may already speculate that we are starting to see a consequence of local 
IR-luminous galaxies often not having star formation spread over a large disk.
If reaching high IR luminosities requires compressing the limited gas content 
of local galaxies anyway, for example by interactions or mergers, then 
favourable conditions for high BHAR will be granted automatically as a 
side effect. This agrees with the well known frequent presence of AGN in 
IR luminous galaxies \citep[e.g.,][]{lutz98,veilleux95,veilleux99,veilleux09}.

\section{Conclusions}

We have analyzed the half light radius of far-infrared 70~\mum\ emission
\reblue\ 
in the hosts of 277 z$<$0.06 SWIFT-BAT selected AGN and 515 z$<$0.06 non-BAT
comparison galaxies, using two-dimensional gaussian fits and subtraction
of the PSF width in quadrature. We find:
\begin{itemize}
\item For both AGN and comparison galaxies, there is a large $\sim$2~dex 
size scatter,
reflecting the wide range of star formation distributions in local galaxies.
\item At modest log(\lfir\ [\lsun\/]) 8.5--10.5 (SFR$\lesssim$6~\msunyr\/), 
the median FIR size of AGN hosts is \reblue\/$\lesssim$1~kpc, a factor
about 2 smaller than the non-BAT comparisons. No such difference is observed
at higher far-infrared luminosities.
\item This size deficit is mostly caused by a more compact distribution of
star formation and gas in the AGN hosts, but a contribution of compact 
AGN-heated dust cannot be excluded for some objects with extreme 
\lbolagn\//\lfir\/.
\item In the context of AGN--host coevolution where SFR and BHAR are fed from
the same general gas supply, these findings argue for a more compact 
SFR and gas distribution favouring AGN feeding. Large scatter remains in 
this link 
because of the large $\sim$kpc scale probed by the SFR data, and the 
possibility of rapid accretion variations that smear the distinction of
AGN and comparison galaxies.
\item The lack of a size difference between AGN hosts and comparison galaxies 
at higher \lfir\ 
may relate to such infrared luminosities mostly requiring compact star 
formation in the first place. 
\end{itemize}

\begin{acknowledgements}
We thank the referee for comments that helped improving the paper.
PACS has been developed by a consortium of institutes led by MPE
(Germany) and including UVIE (Austria); KUL, CSL, IMEC (Belgium); CEA,
OAMP (France); MPIA (Germany); IFSI, OAP/OAT, OAA/CAISMI, LENS, SISSA
(Italy); IAC (Spain). This development has been supported by the funding
agencies BMVIT (Austria), ESA-PRODEX (Belgium), CEA/CNES (France),
DLR (Germany), ASI (Italy), and CICYT/MCYT (Spain). This work has made use of
the NASA/IPAC Extragalactic Database (NED), which is operated by the Jet
Propulsion Laboratory, California Institute of Technology, under contract 
with the National Aeronautics and Space Administration. This research has 
made use of the NASA/IPAC Infrared Science Archive, which is operated by 
the Jet Propulsion Laboratory, California Institute of Technology, under 
contract with the National Aeronautics and Space Administration.
\end{acknowledgements}

\bibliographystyle{aa}
\bibliography{31423}

\begin{appendix}
\section{Far-infrared sizes for BAT sample and reference 
sample}
Table~\ref{tab:data} lists the derived 70~\mum\ sizes and other 
quantities for our samples.
A portion is shown here for guidance, the full table is available
electronically at the CDS VizieR service.
\label{app:datatable}

\begin{table}[h]
\caption{BAT and reference samples}
\begin{tabular}{lrrcrcrlrr}\hline
Name& RA  &DEC  &Sample&Scale      &\multicolumn{2}{c}{\reblue}&Band&log(\lfir\/)&log(\lbat\/)\\
    &J2000&J2000&      &kpc/\arcsec&\multicolumn{2}{c}{kpc}    &    &\lsun       &erg/s\\
 (1)&(2)&(3)&(4)&(5)&\multicolumn{2}{c}{(6)}&(7)&(8)&(9)\\ \hline
UGC 12914           &   0.4110&  23.4825& Ref& 0.297&   5.876&$\pm$0.032&  70& 10.16&      \\
UGC 12915           &   0.4256&  23.4956& Ref& 0.295&   2.086&$\pm$0.003&  70& 10.49&      \\
PG 0003+199         &   1.5805&  20.2027& BAT& 0.519&   0.399&$\pm$0.075&  70&  9.77& 43.45\\
NGC 23              &   2.4716&  25.9237& Ref& 0.310&   1.004&$\pm$0.003&  70& 10.75&      \\
NGC 34              &   2.7780& -12.1076& Ref& 0.398&   0.253&$\pm$0.017&  70& 11.28&      \\
Arp 256N            &   4.7081& -10.3604& Ref& 0.549&   3.159&$\pm$0.016&  70& 10.34&      \\
Arp 256S            &   4.7122& -10.3771& Ref& 0.544&   0.876&$\pm$0.010&  70& 11.13&      \\
2MASX J00253292+6821&   6.3847&  68.3625& BAT& 0.246&   0.307&$\pm$0.050&  70&  9.07& 42.77\\
CGCG 535-012        &   9.0875&  45.6649& BAT& 0.934&   1.909&$\pm$0.287&  70& 10.03& 43.92\\
$\ldots$ \\ \hline
\end{tabular}
\tablefoot{\\
(1) Source name.\\
(2),(3) Far-infrared source position, as measured from the gaussian fits.\\
(4) Sample --  BAT or reference (Ref).\\
(5) Scale for the adopted distance and cosmology.\\
(6) Far-infrared half light radii at 70~\mum\/. Half
light radii are based on subtracting in quadrature observed width and PSF 
width. The errors are only statistical and do not include systematics due to 
the non-gaussianity of the real source structure and the PSF.\\
(7) Band used to derive \reblue\/. \reblue\ is derived from Herschel-PACS 
70~\mum\ data, with few exceptions where it was scaled from \regreen\ 
(measured at 100~\mum\/), using \reblue\/ $=0.85\times$\regreen\/. These are 
marked 100 in the Band column.\\
(8) 40-120~\mum\ far-infrared luminosity.\\
(9) \swift\/-BAT extremely hard X-ray luminosity.
}
\label{tab:data}
\end{table}

\end{appendix}

\end{document}